\documentclass[aps,prb,showpacs,reprint,superscriptaddress]{revtex4-1}

\usepackage{graphicx}
\usepackage{color}
\usepackage{amsmath}
\usepackage{bbm}
\usepackage{amssymb}

\usepackage{dsfont}

\usepackage[utf8]{inputenc}	
\usepackage[T1]{fontenc}

\input epsf

\begin{document}

\title{Superconducting dome in the LaAlO$_3$/SrTiO$_3$ interfaces as a direct effect of the extended s-wave symmetry of the gap}

\author{M. Zegrodnik}
\email{michal.zegrodnik@agh.edu.pl}
\affiliation{Academic Centre for Materials and Nanotechnology, AGH University of Science and Technology, Al. Mickiewicza 30, 30-059 Krakow,
Poland}
\author{P. W\'ojcik}
\email{pawel.wojcik@fis.agh.edu.pl}
\affiliation{AGH University of Science and Technology, Faculty of Physics and Applied Computer cience, Al. Mickiewicza 30, 30-059 Krakow, Poland}

\date{02.05.2020}

\begin{abstract}
The two-dimensional electron gas (2DEG) at the LaAlO$_3$/SrTiO$_3$ interface exhibits gate tunable superconductivity with a domelike shape of $T_{\rm{C}}$ as a function of electron concentration. Here, we propose that the experimentally observed behavior can be explained as a direct effect of the dominant $extended$ $s$-$wave$ symmetry of the superconducting gap. Our results agree very well with the experimental data. As shown, neither the correlation effects nor the spin-orbit coupling influence significantly the physical picture of the paired state steaming out from our analysis.
\end{abstract}


\maketitle
\section{Introduction}
The two-dimensional electron gas (2DEG) at the interface between LaAlO$_3$ and SrTiO$_3$ (LAO/STO) has attracted growing interest as a fundamental system to study the interplay between superconductivity, spin-orbit interaction, and magnetism. It has been well established that LAO/STO exhibits gate tunable superconductivity with the dome-like shape of $T_{\rm{C}}$ as a function of gate voltage \cite{Reyren1196,Rout2017,joshua2012universal,maniv2015strong,Biscaras,biscaras2010two,caviglia2008electric}. The origin of such behavior still remains unclear and is the subject of an ongoing debate\cite{Gorkov2016,Edge2015,Ruhman2016,Gariglio2016}, which mainly concentrates around the 
role of electronic correlations\cite{maniv2015strong,Monteiro2019}, multiband effects~\cite{Trevisan2018, Wojcik2020}, and spin-orbit interaction\cite{Khalsa2013,Diez2015,Zhong2013,Shalom2010,Rout2017}.

It is believed that the low-energy electronic structure of LAO/STO interface comes from the $t_{2g}=\{d_{xy},d_{xz},d_{yz}\}$ orbitals of the Ti atoms\cite{Khalsa2013}. According to the correlation effect scenario, the inter-orbital Coulomb repulsion, leads to nonmonotonic population of the low energy $xy$ mobile band resulting in the dome-like shape of $T_{\rm{C}}$ as the electron concentration increases\cite{maniv2015strong}. Within such approach the $xy$ band contributes significantly to the formation of the SC state and the maximal $T_{\rm{C}}$ appears close to the Lifshitz transition (LT) where the change of electron concentration monotonicity takes place\cite{Smink}. Other reports suggest that the suppression of $T_{\rm{C}}$ above LT is due to a strong pair-breaking effect resulting from a repulsive interband interaction\cite{Trevisan2018, singh2018gap}. The negative influence of LT on the pairing in LAO/STO has been investigated experimentally in Refs. \onlinecite{joshua2012universal,singh2018gap}. On the other hand, according to standard BCS theory an enhancement of $T_{\rm{C}}$ should appear after new states are available for the Cooper pair formation above the Lifshitz transition\cite{Shanenko2015,Wojcik2014_pss}. This discrepancy has not been completely resolved so far. Additionally, in contradiction to the mentioned proposals\cite{maniv2015strong,Trevisan2018, singh2018gap} some of the experimental analysis show that the upper $xz/yz$ bands play the major role in the formation of the paired state and LT does not correspond to maximal $T_{\rm{C}}$. Instead, the superconductivity sets in close to the point where the multiband behavior appears (at the Lifshitz transition)\cite{Gariglio2016,Biscaras}.

Finally, one should note that using magnetotransport measurements it has been established that the spin-orbit couplinig (SOC) energy follows the nonmonotonic dependence of $T_{\rm{C}}$\cite{Shalom2010, Rout2017, Yin2019}, what may indicate that SOC constitutes a significant factor, which tunes the pairing strength leading to the dome-like behavior of $T_{\rm{C}}$. However, for some of the interface orientation such effect has not been reported\cite{singh2018gap}. Also, the inteplay between superconductivity and SOC has not been recognised in detail so far. Therefore, it is not clear if the spin-orbit energy is in fact the primary cause of the characteristic shape of the phase diagram or a secondary effect.

In spite of several mentioned proposals aiming in explanation of the SC dome in LAO/STO, the satisfactory theoretical reconstruction of T$_C$ as a function of gate voltage has not been reached so far.

Here, we show that the appearance of the superconducting dome as a function of gate voltage in the LAO/STO interfaces can be explained as a sole result of the $extended$ $s$-$wave$ superconducting gap symmetry appearing in the range of relatively low electron concentrations. Our approach leads to a very good agreement with the available experimental data of the gate voltage dependance of $T_{\rm{C}}$. To analyze the influence of electronic correlations we carry out calculations with the inclusion of the Coulomb repulsion terms by using both the Hartree-Fock method (HF) and statistically consistent Gutzwiller approximation (SGA)\cite{Gutzwiller1963,Jedrak2011}. We also provide the results obtained in the presence of Rashba and atomic components of the spin-orbit coupling. As shown neither the correlation effects nor the SOC influence significantly the physical picture of the paired state steaming out from our analysis.



\section{Model and method}
We consider the two-dimensional electronic gas at the LAO/STO interface with (001) orientation. The Hamiltonian of the system is taken as
\begin{equation}
    \hat{H}=\hat{H}_{TBA}+\hat{H}_U+\hat{H}_{SC},
    \label{eq:Hamiltonian_general}
\end{equation}
where the subsequent terms correspond to the single-particle part, Coulomb repulsion, and the real-space pairing, respectively. For clarity, in this Section we show the Hartree-Fock-BCS treatment of the considered model without the inclusion of the the spin-orbit coupling (SOC). The form of $\hat{H}_{TBA}$ with the inclusion of SOC is deferred to Appendix A, while the SGA approach as applied to the considered model is provided in Appendix B.

\begin{figure}
 \centering
 \includegraphics[width=0.5\textwidth]{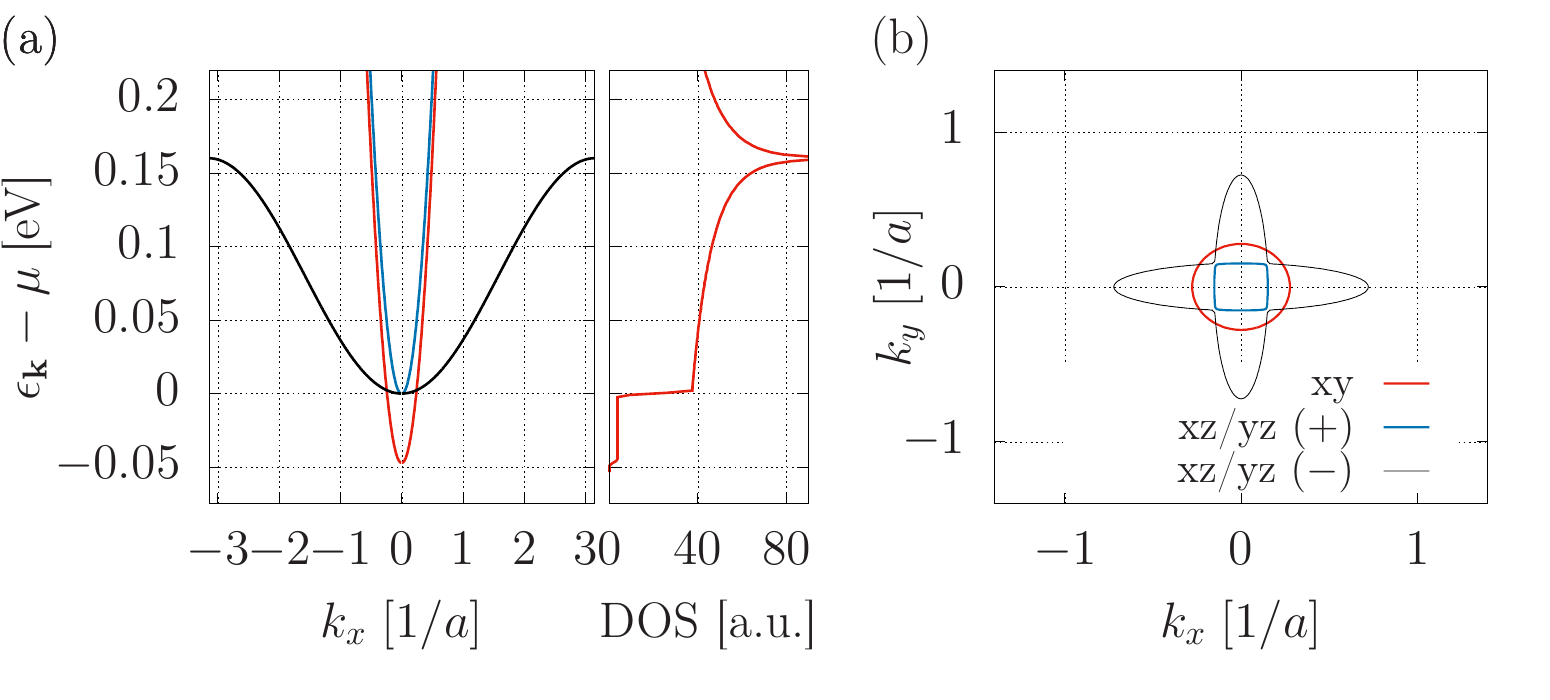}
 \caption{(a) Band structure and the density of states of the three-orbital model representing the 2DEG at the LAO/STO interface. The $xy$-band (red solid line) is $47\;$meV lower in energy at the $\Gamma$ point than the two hybridized $xz/yz$-bands (black and blue solid lines); (b) The Fermi surfaces of the system for $\mu=20\;$meV.}
 \label{fig:band_struct}
\end{figure}

The single particle part of Hamiltonian (\ref{eq:Hamiltonian_general}) is expressed within the three-orbital tight binding approximation\cite{Khalsa2013,maniv2015strong}
\begin{equation}
 \hat{H}_{TBA}=\sum_{\mathbf{k}l\sigma}(\epsilon^{l}_{\mathbf{k}}-\mu)\hat{c}^{\dagger}_{\mathbf{k}l\sigma}\hat{c}_{\mathbf{k}l\sigma}+\sideset{}{''}\sum_{\mathbf{k}ll'\sigma}\epsilon_{h\mathbf{k}}\hat{c}^{\dagger}_{\mathbf{k}l\sigma}\hat{c}_{\mathbf{k}l'\sigma}
 \label{eq:Hamiltonian_start}
 \end{equation}
where $\mu$ is the chemical potential, $\hat{c}^{\dagger}_{\mathbf{k}l\sigma}$ ($\hat{c}_{\mathbf{k}l\sigma}$) are the creation (anihilation) operators of electrons with momentum $\mathbf{k}$, spin $\sigma$, and orbital index $l=xy,\;xz,\;yz$ corresponding to $d_{xy}$, $d_{xz}$, $d_{yz}$ orbitals of the Ti atoms placed on a square lattice. The double primmed summation in the second term is restricted to the two upper $xz$ and $yz$ bands only, with $l\neq l'$. The bare (unhybridized) dispersion relations have the form
\begin{equation}
\begin{split}
    \epsilon^{xy}_{\mathbf{k}}&=4t_l-\Delta_E-2t_l\cos{k_x}-2t_l\cos{k_y},\\
    \epsilon^{xz}_{\mathbf{k}}&=2t_l+2t_h-2t_l\cos{k_x}-2t_h\cos{k_y},\\
    \epsilon^{yz}_{\mathbf{k}}&=2t_l+2t_h-2t_h\cos{k_x}-2t_l\cos{k_y},
\end{split}
\end{equation}
and the mixing between the $xz$- and $yz$-bands is the following
\begin{equation}
    \epsilon_{h\mathbf{k}}=2t_d\sin{k_x}\sin{k_y},
\end{equation}
where the tight-binding parameters have been taken from Ref. \onlinecite{maniv2015strong} and are: $t_l=875\;$meV, $t_h=40\;$meV, $t_d=40\;$meV, $\Delta_E=47\;$meV. The resulting band structure of the model is presented in Fig. \ref{fig:band_struct}, and consists of the $xy$-band which is lower in energy by $\Delta_E$ at the $\Gamma$ point than the two hybridized $xz/yz$ bands. It is worth noting that the density of states in the bottom $xy$-band is significantly smaller than in the two upper $xz/yz$ bands (cf. Fig. \ref{fig:band_struct}).\\
The second term of our model Hamiltonian (\ref{eq:Hamiltonian_general}) has the form
\begin{equation}
 \hat{H}_{U}=U\sum_{il}\hat{n}_{il\uparrow}\hat{n}_{il\downarrow}+V\sideset{}{'}\sum_{ill'}\hat{n}_{il}\hat{n}_{il'},
 \label{eq:Hamiltonian_U}
\end{equation}
where $U$ and $V$ are the intra- and inter-orbital Coulomb repulsion integrals and the primmed sumation is restricted to $l\neq l'$. For simplicity we take $U=V\equiv2\;$eV, which corresponds to the value calculated in Ref. \onlinecite{Breitschaft}.

In our approach the superconducting state is introduced by a real-space intersite intraorbital pairing as well as the interorbital pair hopping terms
\begin{equation}
\begin{split}
 \hat{H}_{SC}&=-J\sum_{ijl}\hat{c}^{\dagger}_{il\uparrow}\hat{c}^{\dagger}_{jl\downarrow}\hat{c}_{il\downarrow}\hat{c}_{jl\uparrow} -J^{\prime}\sideset{}{'}\sum_{ijll'}\hat{c}^{\dagger}_{il\uparrow}\hat{c}^{\dagger}_{jl\downarrow}\hat{c}_{il'\downarrow}\hat{c}_{jl'\uparrow},
 \end{split}
 \label{eq:Hamiltonian_pairing}
\end{equation}
for which the interorbital pair hopping energy, $J^{\prime}$, is one order of magnitude smaller than the intraorbital coupling constant, $J$.  

After the application of the standard Hartree-Fock-BCS treatment of the pairing part and the Coulomb interaction terms, we express the model Hamiltonian in the following form in reciprocal space
\begin{equation}
\begin{split}
 \hat{H}&=\sum_{\mathbf{k}}\mathbf{\hat{f}}^{\dagger}
_{\mathbf{k}}\mathbf{\hat{H}}_{\mathbf{k}}\mathbf{\hat{f}}_{\mathbf{k}}+\sum_{\mathbf{k}l}\xi^l_{\mathbf{k}}-N\sum_{l}\bigg(U\frac{n^2_{l}}{4}+V\sum_{l'(l'\neq l)}n_{l}n_{l'}\bigg)\\
&+16\frac{J^2-(J')^2}{(J-J')^2(J+2J')}N\sum_l\bigg((\Delta^s_l)^2+(\Delta^d_l)^2\bigg)\\
&+16\frac{(J')^2-J'J}{(J-J')^2(J+2J')}N\sum_{ll'(l\neq l')}\bigg(\Delta^s_l\Delta^s_{l'}+\Delta^d_l\Delta^d_{l'}\bigg),
\end{split}
\label{eq:ham_HF}
\end{equation}
where $N$ is the number of atomic sites, $n_l=\langle\hat{n}_{il\uparrow}\rangle+\langle\hat{n}_{il\downarrow}\rangle$, while $\Delta^s_l$ and $\Delta^d_l$ are the $extended$ $s$- and $d$-$wave$ pairing amplitudes
\begin{equation}
    \Delta_l^s=\frac{1}{4}\sum_{j(i)}\gamma^s_{ij}\Delta_{ijl},\quad     \Delta_l^d=\frac{1}{4}\sum_{j(i)}\gamma^d_{ij}\Delta_{ijl}.
    \label{eq:Delta_amplitudes}
\end{equation}
The summations above run over the four nearest-neighbor atomic sites of $\mathbf{R}_i$. These sums do not depend on the position of $\mathbf{R}_i$ since the system is homogeneous. The $extended$ $s$-$wave$ and $d$-$wave$ real-space symmetry factors are $\gamma^s_{i,j}\equiv1$ and $\gamma^d_{i,j}=1$ ($\gamma^d_{i,j}=-1$) for $\mathbf{R}_j=\mathbf{R}_i\pm\hat{x}$ ($\mathbf{R}_j=\mathbf{R}_i\pm\hat{y}$), while the $\Delta_{ijl}$ parameters correspond to the combination of the anomalous superconducting expectation values   
\begin{equation}
   \Delta_{ijl}=-J\langle\hat{c}^{\dagger}_{il\uparrow}\hat{c}^{\dagger}_{jl\downarrow}\rangle-J^{\prime}\sum_{l'(l'\neq l)}\langle\hat{c}^{\dagger}_{il'\uparrow}\hat{c}^{\dagger}_{jl'\downarrow}\rangle.
   \label{eq:Delta_real_space}
\end{equation}
As one can see due to the pair-hopping terms there is a small contribution to the pairing amplitude between particular orbitals which comes from the remaining orbitals of the model (the second term above). Such mechanism connects all the SC amplitudes and guarantees the appearance of a single critical temperature. \newline
In Eq. (\ref{eq:ham_HF}) we have introduced the six-component composite operator
\begin{equation}
 \mathbf{\hat{f}}^{\dagger}_{\mathbf{k}}\equiv(\hat{c}^{\dagger}_{\mathbf
{k},xy\uparrow},\hat{c}_{-\mathbf{k},xy\downarrow},\hat{c}^{\dagger}_{\mathbf
{k},xz\uparrow},\hat{c}_{-\mathbf{k},xz\downarrow},\hat{c}^{\dagger}_{\mathbf
{k},yz\uparrow},\hat{c}_{-\mathbf{k},yz\downarrow})\;,
\end{equation}
while the form of the 6$\times$6 matrix Hamiltonian is
\begin{equation}
\mathbf{\hat{H}}_{\mathbf{k}}=\left(\begin{array}{cccccc}
 \xi^{xy}_{\mathbf{k}} & \Delta^{xy}_{\mathbf{k}} & 0 & 0 & 0 & 0\\
\Delta^{xy}_{\mathbf{k}} & -\xi^{xy}_{\mathbf{k}} & 0 & 0 & 0 & 0\\
0 & 0 & \xi^{xz}_{\mathbf{k}} & \Delta^{xz}_{\mathbf{k}} & \epsilon_{h\mathbf{k}} & 0\\
0 & 0 & \Delta^{xz}_{\mathbf{k}} & -\xi^{xz}_{\mathbf{k}} & 0 & -\epsilon_{h\mathbf{k}}\\
0 & 0 & \epsilon_{h\mathbf{k}} & 0 & \xi^{yz}_{\mathbf{k}} & \Delta^{yz}_{\mathbf{k}}\\
0 & 0 & 0 & -\epsilon_{h\mathbf{k}} & \Delta^{yz}_{\mathbf{k}} & -\xi^{yz}_{\mathbf{k}} \\
\end{array} \right)\;.
\label{eq:matrix_H}
\end{equation}
The diagonal elements in Eq. (\ref{eq:matrix_H}) contain the chemical potential and the effective shift of the atomic energy which originates from the Hartree-Fock approximation of the Coulomb interaction terms
\begin{equation}
    \xi^{l}_{\mathbf{k}}=\epsilon^l_{\mathbf{k}}+U\frac{n_{l}}{2}+V\sum_{l'(l'\neq l)}n_{l'}-\mu.
    \label{eq:diagonal_dissp}
\end{equation}
The $\mathbf{k}$-dependent SC gaps appearing in the Hamiltonian matrix (\ref{eq:matrix_H}) can be written in the form
\begin{equation}
    \Delta^l_{\mathbf{k}}=4\Delta^s_l\gamma^s_{\mathbf{k}}+4\Delta^d_l\gamma^d_{\mathbf{k}},
\end{equation}
where the $extended$ $s$- and $d$-$wave$ symmetry factors in the reciprocal space are given below
\begin{equation}
    \gamma^s_{\mathbf{k}}=(\cos k_x  + \cos k_y)/2, \quad \gamma^d_{\mathbf{k}}=(\cos k_x  - \cos k_y)/2.
    \label{eq:symmetry_factors}
\end{equation}
The appearance of both $extended$ $s$- and $d$-$wave$ components in the SC gaps for the $xz$ and $yz$ bare (unhybridized) bands is the consequence of the C$_4$ symmetry breaking in both $d_{xz}$ and $d_{yz}$ orbitals. In the $xy$-band which is C$_4$ symmetric, only one of the mentioned components can appear and the remaining one needs to be zero. According to our calculations only the pure $extended$ $s$-$wave$ pairing appears in that band for the parameter range significant for the analyzed system. Therefore, we obtain
\begin{equation}
    \Delta^{xy}_{\mathbf{k}}=4\Delta^s_{xy}\gamma^s_{\mathbf{k}}.
    \label{eq:delta_k0}
\end{equation}
By carrying out the diagonalization procedure of the $xz/yz$-mixing part of our Hamiltonian one can show that in the resulting hybridized $xz/yz$ bands the $C_4$ symmetry is restored both in the disspersion relations
\begin{equation}
    \epsilon^{xz/yz}_{\mathbf{k}}=\frac{1}{2}\bigg((\epsilon^{xz}_{\mathbf{k}}+\epsilon^{yz}_{\mathbf{k}})\mp\sqrt{(\epsilon^{xz}_{\mathbf{k}}-\epsilon^{yz}_{\mathbf{k}})^2+4\epsilon_{h\mathbf{k}}^2}\bigg),
    \label{eq:epsilon_k12}
\end{equation}
and in the corresponding $\mathbf{k}$-dependent SC gaps
\begin{equation}
    \Delta^{xz/yz}_{\mathbf{k}}=4\Delta_{xz/yz}^s\gamma^s_{\mathbf{k}}\pm4\Delta_{xz/yz}^d\alpha_{\mathbf{k}}\gamma^d_{\mathbf{k}},
    \label{eq:delta_k12}
\end{equation}
where $\Delta_{xz/yz}^s\equiv\Delta_{xz}^s=\Delta_{yz}^s,\; \Delta_{xz/yz}^d\equiv\Delta_{xz}^d=-\Delta_{yz}^d$, while the $\alpha_{\mathbf{k}}$ factor results directly from the hybridization between the $d_{xz}$ and $d_{yz}$ bands 
\begin{equation}
    \alpha_{\mathbf{k}}=\frac{\epsilon^{xz}_{\mathbf{k}}-\epsilon^{yz}_{\mathbf{k}}}{\sqrt{(\epsilon^{xz}_{\mathbf{k}}-\epsilon^{yz}_{\mathbf{k}})^2+4\epsilon_{h\mathbf{k}}^2}}.
\end{equation}
One can check, that by carrying out $\pi/2$-rotations in reciprocal space for Eqs. (\ref{eq:epsilon_k12}) and (\ref{eq:delta_k12}) the same formulas are obtained, meaning that the $C_4$ symmetry is conserved in spite of the fact that both $extended$ $s$- and $d$-$wave$ pairing amplitudes appear.


\begin{figure*}[t!]
 \centering
 \includegraphics[width=1.0\textwidth]{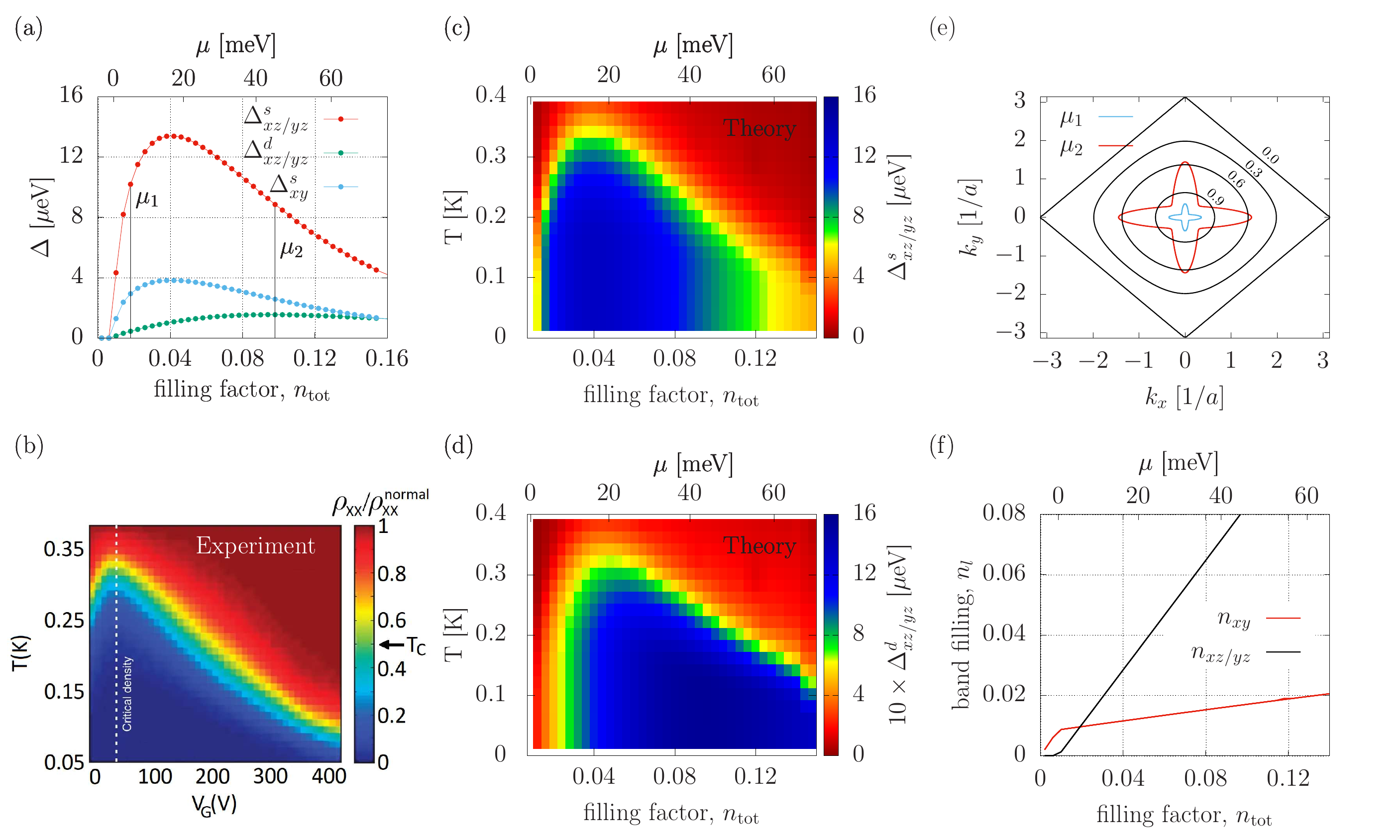}
 \caption{(a) The $extended$ $s$- and $d$-$wave$ pairing amplitudes of the $xy$ band and the two $xz/yz$ hybridized bands [cf. Eqs. (\ref{eq:delta_k0}) and (\ref{eq:delta_k12})] for $T=0\;$K as a function of band filling (bottom axis) and chemical potential (top axis); (b) The experimental phase diagram showing how $T_{\rm{C}}$ changes as a function of gate voltage (taken from Ref. \onlinecite{joshua2012universal}); (c) and (d) The theoretical phase diagrams showing the reconstruction of the dome-like shape of $T_{\rm{C}}$ as a function of electron concentration; (e) Fermi surfaces of the hybridized band corresponding to two exemplary values of the chemical potential, $\mu_1$ and $\mu_2$, marked in (a). The black solid lines in the plot represent the isolines of the $extended$ $s$-$wave$ symmetry factor corresponding to $\gamma_{\mathbf{k}}=0.9,\;0.6,\;0.3,\;0.0$ [cf. Eq. (\ref{eq:symmetry_factors})]; (f) Band filling components corresponding to the $xy$-band ($n_{xy}$) and the two hybridized $xz/yz$-bands ($n_{xz/yz}=n_{xz}+n_{yz}$).}
 \label{fig:phase_diags}
\end{figure*}

As shown above, the Coulomb interaction terms appearing in Eq. (\ref{eq:Hamiltonian_U}) are treated with the use of the Hartree-Fock (HF) approximation leading to an effective shift of atomic energy, dependant on the filling of particular bands [cf. Eq. (\ref{eq:diagonal_dissp})]. Due to the fact that such mean-field procedure neglects most of the electron correlations effects we also apply the statistically consistent Gutzwiller approximation (SGA)\cite{Jedrak2011,Kaczmarczyk2013,Zegrodnik_2013} for comparison. Within the SGA approach apart from the standard mean-field atomic energy shifts, the correlation induced renormalization of both electron hopping and pairing is taken into account (cf. Appendix B).

\section{Results}
We start from the model with no Coulomb repulsion terms included ($U=V=0$) and analyze the superconducting properties of the 2DEG at the LAO/STO interface as a function of the chemical potential, $\mu$, or equivalently the filling factor, $n_{\mathrm{tot}}=\sum_{il\sigma}n_{il\sigma}/N$ ($N$-number of atomic sites). Note that, by increasing the gate voltage in experiments one adds electrons to the system what leads to increase of both $\mu$ and $n_{\mathrm{tot}}$. The effect of Coulomb repulsion is analyzed later on both by the use of HF and SGA approximations. At the end we also show the influence of the SOC terms on our results. In all the calculations the value of coupling constant has been set to $J=0.165\;$eV so as to reproduce the maximal critical temperature $T_C\approx 0.35\;$K which is measured in experiments.

Results for $T=0\;$K presented in Fig. \ref{fig:phase_diags}(a) show that the $extended$ $s$-$wave$ pairing amplitude in the two hybridized bands ($\Delta_{xz/yz}^s$) constitutes the dominant contribution to the superconducting phase and it reproduces the dome-like shape of the critical temperature as a function of gate voltage, which is reported in experiments\cite{joshua2012universal,Bert2012} [cf. Fig. \ref{fig:phase_diags}(b)]. The gap amplitude in the low-energy band ($\Delta_{xy}^s$) follows the trend of the amplitudes in the two upper bands ($\Delta_{xz/yz}^s$). This results from the fact that in the former the density of states is too low for the pairing to appear naturally at the given value of $J$ (cf. Fig. \ref{fig:band_struct}). Therefore, the gap in the $xy$-band is induced by the pair hopping processes from the two upper bands with significantly higher DOS. Note, that the remaining $\Delta_{xz/yz}^d$ gap amplitude has a negligible influence on the SC properties of the system being one order of magnitude smaller than $\Delta_{xz/yz}^s$. 

In Figs. \ref{fig:phase_diags}(c) and (d) we present the results for $T>0\;$K, which show that indeed the dome-like shape of $T_{\rm{C}}$ as a function of the filling factor (and chemical potential) is reproduced in our model and matches very well the experimental data provided for comparison in Fig. \ref{fig:phase_diags}(b). Here, we do not show the gap amplitude in the lower band, $\Delta_{xy}^s$, as it has virtually the same behavior as $\Delta_{xz/yz}^s$. However, the former is scaled down to approximately three times lower values than the latter [cf. Fig. \ref{fig:phase_diags}(a)]. One should note that the fall of both $\Delta^s_{xz/yz}$ and $T_{\rm{C}}$ above the optimal $\mu$, for which maximal values are obtained, is not determined by the structure of density of states since the latter does not show any peak in the corresponding energy range between $0$ and $60\;$meV (cf. Fig. \ref{fig:band_struct}).

The explanation for the obtained dome-like shape of $T_{\rm{C}}$ within our approach is the following. As shown in Figs. \ref{fig:phase_diags}(a), (c), and (d) the $extended$ $s$-$wave$ pairing amplitude dictates the changes of $T_{\rm{C}}$ as the number of electrons increases. For such situation one can distinguish between two regions. The first one corresponds to very low electron concentrations when the Fermi surface is contained in the close proximity of the $\Gamma$ point in the center of the Brillouin zone [cf. Fig. \ref{fig:phase_diags}(e)]. In this regime the $extended$ $s$-$wave$ symmetry factor $\gamma^s_{\mathbf{k}}\approx 1$ at the Fermi surface and it does not tune the value of the gap significantly. As one can see in Fig. \ref{fig:phase_diags}(e) for $\mu$ below the optimal value, the Fermi surface (blue line) is placed inside the closed isoline representing $\gamma^s_{\mathbf{k}}=0.9$. In this regime a standard behavior of rising $T_{\rm{C}}$ with the chemical potential appears, similarly as in the conventional case of constant SC gap ($\Delta_{\mathbf{k}}\equiv\Delta$) within the real-space pairing scenario. However, as $\mu$ increases the Fermi surface expands and moves closer to the nodal lines of the $extended$ $s$-$wave$ symmetry factor, where the gap closes [cf. Fig. \ref{fig:phase_diags}(e)]. As one can see for $\mu$ above the optimal value, the Fermi surface (red line) reaches the isolines corresponding to $\gamma^s_{\mathbf{k}}=0.6$. At this point the suppression of the gap at the Fermi surface resulting from the $extended$ $s$-$wave$ symmetry becomes significant. In this regime superconductivity is gradually weakened as one adds electrons to the system. Between the two regions the optimal chemical potential is placed, for which the maximal $T_{\rm{C}}$ appears. 

In Fig. \ref{fig:phase_diags}(f) we show how the electrons injected into the system are distributed between the $xy$ band (red line) and the two hybridized $xz/yz$ bands (black line). The Lifshitz transition corresponds to $\mu=0$, for which the two hybridized bands begin to be populated and superconductivity sets in [cf. Fig. \ref{fig:phase_diags} (a)]. This result is in agreement with the experimental data presented in Ref. \onlinecite{Biscaras}, where the transition from the single to multiband behavior takes place in close proximity to the transition between the normal and superconducting state. However, in this respect the experimental situation is not completely clear, since in Ref. \onlinecite{joshua2012universal} the authors claim that the maximum $T_{\rm{C}}$ corresponds to the Lifhitz transition, which would be in contradiction both to the data presented in Ref. \onlinecite{Biscaras} and with our results. 

It should be noted that in Fig. \ref{fig:phase_diags}(f) both $n_{xy}$ and $n_{xz/yz}$ are monotonically increasing functions of the total electron concentration. As we show below, the nonmonotonic behavior of $n_{xy}$, which is seen in experiments\cite{Smink}, can be reproduced only after the inclusion of the Coulomb repulsion terms.



\begin{figure}
 \centering
 \includegraphics[width=0.5\textwidth]{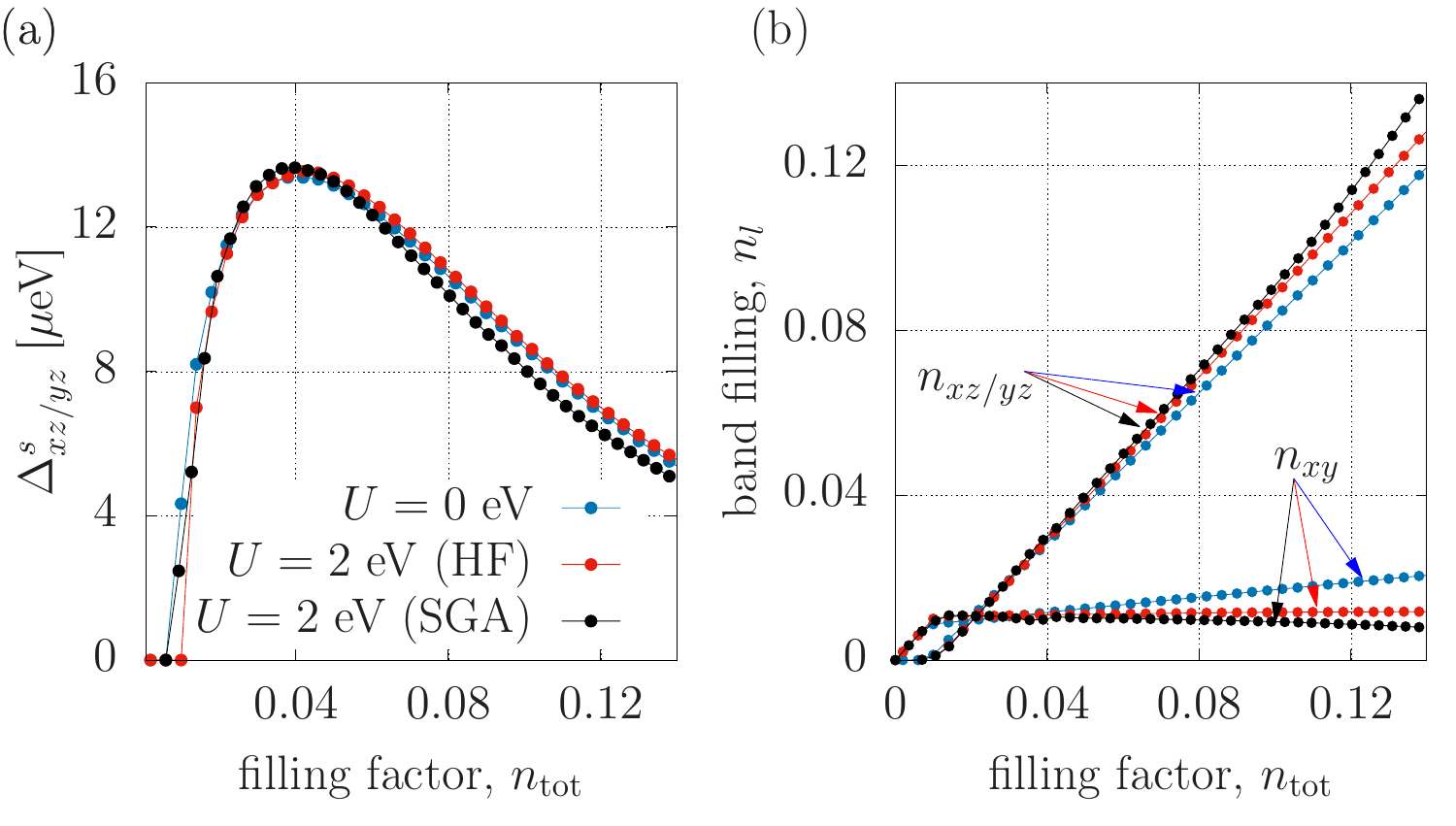}
 \caption{The SC gap amplitude in the hybridized bands as a function of the filling factor for $U=V=0\;$eV and for $U=V=2\;$eV within the HF and SGA approaches. (b) The charge distribution between the bottom and the two hybridized bands of the model as we increase the filling factor for the same model parameters and calculation methods as in (a).}
 \label{fig:electron_interactions}
\end{figure}

In Fig. \ref{fig:electron_interactions}(a) we show the $extended$ $s$-$wave$ pairing amplitude $\Delta_{xz/yz}^s$ as a function of the filling factor for the case of nonzero Coulomb interaction integrals $U$ and $V$. As one can see, both HF and SGA methods lead to very similar results, which additionally are very close to those obtained earlier for the case of no Coulomb interactions ($U=V=0$). Therefore, one can conclude that the interactions do not influence significantly the considered here paired state in the parameter regime significant for the LAO/STO interfaces. However, the interorbital Coulomb term makes the carrier density in the low-energy $xy$-band a nonmonotonic function of band filling in agreement with the experimental data presented in Ref. \onlinecite{Smink} (Fig. 3 in that paper). The influence of that mechanism on superconductivity, which was proposed in Ref. \onlinecite{maniv2015strong}, does not play a role here because the pairing in our model originates mainly from the two upper $xz/yz$ bands and not from the bottom $xy$-band.

One should also note that for relatively small number of electrons in the system, the number of multiple occupancies on a single atomic site leading to interactions is very low. The regime analyzed here corresponds to $\lesssim 0.1$ electron per lattice site and is far form $n_{\mathrm{tot}}\approx 1, 2, 3, ...$ for which the electron-electron correlations are enhanced. In such case, the  correlation effects taken into account within the SGA method are suppressed making the HF and SGA results very close.

\begin{figure}
 \centering
 \includegraphics[width=0.45\textwidth]{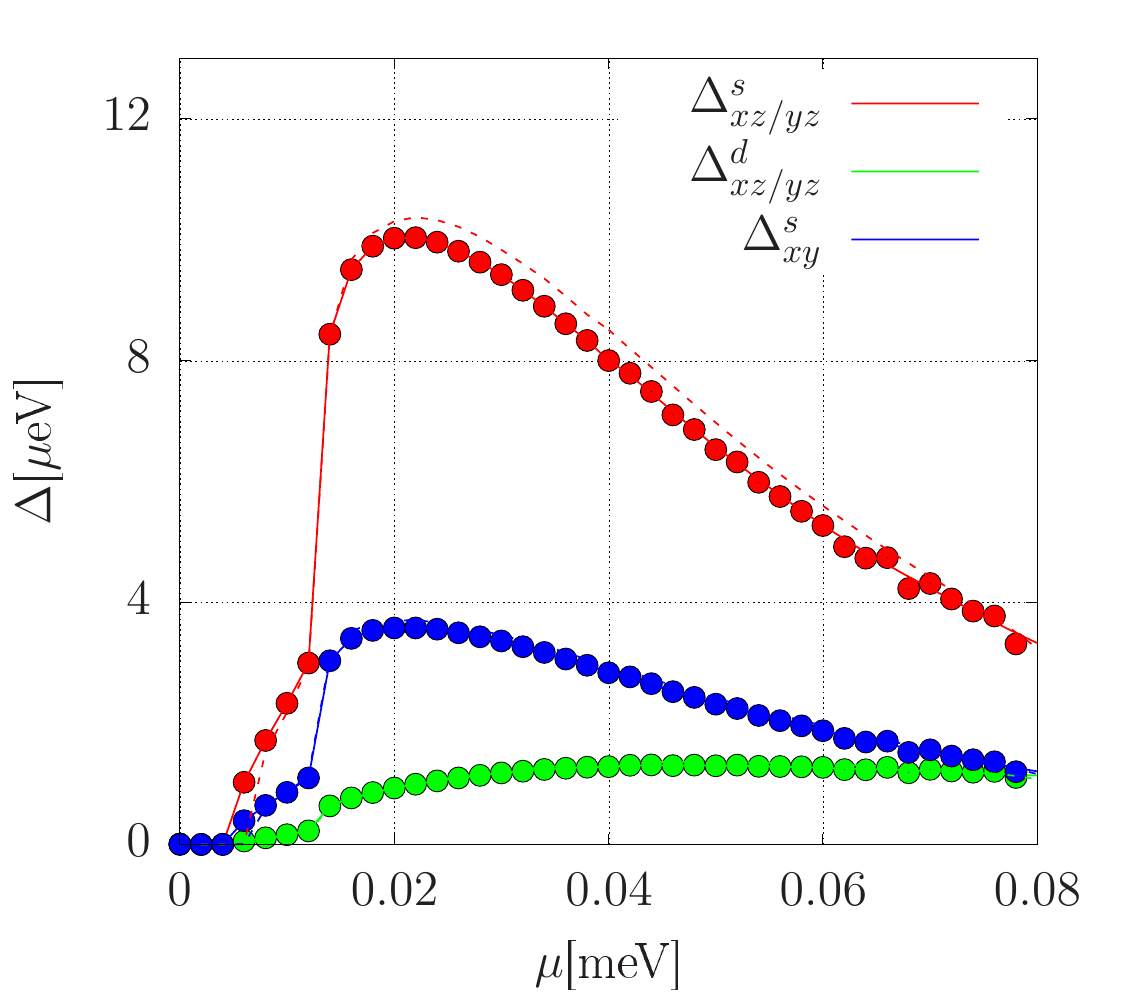}
 \caption{The SC gap amplitude in the hybridized bands as a function of the chemical potential. The results have been obtained with the inclusion of 
the atomic ($\Delta_{SO}=30\;$meV) and Rashba ($\Delta_{RSO}=30\;$meV) SOC terms. Dashed lines correspond to calculations with the atomic SOC 
component only.}
 \label{fig:SOC_terms}
\end{figure}

For the sake of completeness, in Fig. \ref{fig:SOC_terms} we also show the data obtained with the inclusion of both Rashba and atomic SOC terms (cf. Appendix A) with the energies set to $\Delta_{RSO}=30\;$meV and $\Delta_{SO}=30\;$meV, respectively.
As one can see the SC gap amplitudes are slightly suppressed by SOC, as well as a small anomaly appears for very low electron concentrations. The latter is caused by the mixing between the bottom $xy$-band and the two upper $xz/yz$-bands, introduced by the SOC, which changes the density of states in that electron concentration region (cf. Fig. 3 from Ref. \onlinecite{joshua2012universal}). Nevertheless, the dome-like behavior shown in Fig.~\ref{fig:phase_diags}(a) remains almost unchanged.

\section{Conclusion}
As we have shown the appearance of the superconducting dome in the LAO/STO interfaces can be explained as a sole result of the $extended$ $s$-$wave$ symmetry of the gap, which appears in the intersite real-space pairing scenario. The mechanism leading to the SC dome reconstruction is based on a simple fact that the $\mathbf{k}$-dependence of the gap results in a significant suppression of the pairing, but only when the Fermi surface is placed relatively far from the $\Gamma$ point in the Brillouin zone. Our theoretical results are in very good agreement with the available experimental data. To the best of our knowledge such high degree of reconstruction of the T$_C$ dome has not been obtained so far within any other theoretical proposal. It should be noted that, in our approach neither the spin-orbit coupling nor the electron correlation effects are responsible for the SC dome appearance. Furthermore, the calculations carried out with the inclusion of the Coulomb repulsion (by using HF and SGA methods) as well as SOC (Rashba and atomic) terms show that the two factors does not influence significantly the obtained phase diagram of the LAO/STO interface. 

An important question concerns the origin of the considered here pairing mechanism described by Eq. (\ref{eq:Hamiltonian_pairing}). In this respect, an interesting theoretical proposal which can be related with the scenario analyzed by us has been provided very recently in Ref. \onlinecite{Pekker2020}. According to this concept electron pairing can be mediated by the ferroelastic domain walls which are ubiquitous at the LAO/STO interface. Orientation of these domains is known to couple to the electron density leading to the alternatively occurring electron-rich and electron-poor regions. As shown, the ferroelastic domains support low-energy excitation at the LAO/STO interface, resulting in superconductivity around the edges of electron-rich regions. Such mechanism leads to a real-space intersite pairing mechanism which can stabilize an $extended$ $s$-$wave$ superconducting state similar as in our case.

Even more promising mechanism, which may be compatible with the paired state considered here is based on the fluctuations of momentum-based multipoles as analyzed in Ref.~\onlinecite{Sumita2020}. Within such concept the interaction vertex under the crystal symmetry corresponding to STO reveals the $extended$ $s$-$wave$ symmetry of the electron pairing in accordance with our assumption. 

It should be noted, that for particular pairing mechanisms that are discussed for the LAO/STO interfaces the strength of the Cooper pair coupling could depend on the carrier concentration which would modify the resulting structure of the phase diagram. However, a detailed analysis of such situation is beyond the scope of this paper. Also, further experimental exploration needs to be carried out to probe the pairing symmetry and determine the origin of the electron pairing mechanism.


\section{Acknowledgement}
This work was supported by National Science Centre, Poland (NCN) according to decision 2017/26/D/ST3/00109 and in part by PL-Grid Infrastructure.

\appendix
\section{Single-particle Hamiltonian with the inclusion of the spin-orbit interaction}
The calculations of the superconducting gaps with the inclusion of SOC follows the same Hartee-Fock mean field procedure as described in Sec. II. The only difference is the kinetic term in Eq.~(\ref{eq:Hamiltonian_general}) which additionally includes the SOC terms. For the 
theoretical description we use a three-band model of the $t_{2g}$ conduction electrons  with Hamiltonian (expressed in the reciprocal space) given by
\begin{equation}
 \hat{H}_{TBA}=\sum _{\mathbf{k},l,\sigma} \hat{c}^{\dagger}_{\mathbf{k},l,\sigma} \left ( \hat{H}_0 + \hat{H}_{SO} + \hat{H}_{RSO}\right ) 
\hat{c}_{\mathbf{k},l,\sigma},
\end{equation}
where $\hat{c}^{\dagger}_{\mathbf{k},l,\sigma}(\hat{c}_{\mathbf{k},l,\sigma})$ creates (anihilates) electrons of spin $\sigma$ and momentum 
$\mathbf{k}$ in orbitals $l=xy,xz,yz$ and $H_0$ is the tree-band Hamiltonian
\begin{equation}
\hat{H}_{0}=
\left(
\begin{array}{ccc}
 \xi^{xy}_{\mathbf{k}} & 0 & 0\\
 0 & \xi^{xz}_{\mathbf{k}} &  \epsilon_{h\mathbf{k}} \\
 0 & \epsilon_{h\mathbf{k}}  & \xi^{yz}_{\mathbf{k}}
\end{array} \right) \otimes \hat {\sigma} _0\;,
\end{equation}
where the diagonal elements are defined in Eq. (\ref{eq:diagonal_dissp}) while $\hat{\sigma} _{x,y,z}$ and $\hat{\sigma} _{0}$ denote the Pauli matrices and the identity matrix acting on the electron spin. \\
The Rashba spin-orbit $H_{RSO}$ results from the intrinsic electric field at the interface which breaks the inversion symmetry.
The Hamiltonian $\hat{H}_{RSO}$ is given by~\cite{Khalsa2013}
\begin{equation}
\hat{H}_{RSO}= \Delta_{RSO}
\left(
\begin{array}{ccc}
0 & i \sin{k_y} & i \sin{k_x}\\
-i \sin{k_y} & 0 & 0 \\
-i \sin{k_x} & 0 & 0
\end{array} \right) \otimes \hat {\sigma} _0\;,
\end{equation}
where $\Delta _{RSO}$ is the energy of the Rashba SOC. \\
Another source of SOC results from the atomic positions. The SOC related with this kind of asymmetry (atomic-like SOC) is described by the 
Hamiltonian~\cite{Khalsa2013}
\begin{equation}
\hat{H}_{SO}= \frac{\Delta_{SO}}{3}
\left(
\begin{array}{ccc}
0 & i \hat{\sigma _x} & -i \hat{\sigma _y}\\
-i \hat{\sigma _x} & 0 & i \hat{\sigma _z} \\
i \hat{\sigma _y} & -i \hat{\sigma _z} & 0
\end{array} \right) \;,
\end{equation}
where $\Delta _{SO}$ determines the atomic-like spin-orbit energy.

\section{Statistically consistant Gutzwiller approach to the three band model of the LAO/STO interface}
Since the Hartree-Fock approximation neglects most of the correlation effects resulting from a significant magnitude of the Coulomb repulsion we additionally carry out calculations with the use of Statistically Consistent Gutzwiller approximation (SGA)\cite{Jedrak2011,Kaczmarczyk2013,Zegrodnik_2013} dedicated for the strongly correlated electron system. We start from the Gutzwiller-type projected many particle wave function of the form
\begin{equation}
 |\Psi_G\rangle\equiv\hat{P}|\Psi_0\rangle=\prod_{il}\hat{P}_{il}|\Psi_0\rangle \;,
 \label{eq:GWF}
\end{equation}
where $|\Psi_0\rangle$ represents the wave function of uncorrelated state with non-zero anomalous superconducting expectation values and the projection operator $\hat{P}_{il}$ has the form
\begin{equation}
 \hat{P}_{il}\equiv \sum_{\Gamma}\lambda_{\Gamma|il}|\Gamma\rangle_{il\;il}\langle\Gamma|\;,
 \label{eq:P_Gamma}
\end{equation}
where $\lambda_{\Gamma|il}$ are the variational parameters determining relative weights corresponding to $|\Gamma\rangle_{il}$, representing the states from the local basis
\begin{equation}
|\Gamma\rangle_{il}\in \{|\varnothing\rangle_{il}, |\uparrow\rangle_{il}, |\downarrow\rangle_{il},
|\uparrow\downarrow\rangle_{il}\}\;.
\end{equation}
which correspond to empty, singly, and doubly occupied states on the atomic sites with the three types of orbitals ($l\in\{d_{xy},d_{xz},d_{yz}\}$). By minimizing the energy of the system over the variational parameters one reduces the number of configurations which correspond to increased interaction energies, thus, taking into account the many-body correlation effects.

Eq. (\ref{eq:P_Gamma}) represents the general form of the correlation operator. Particularly useful is the form with the following constraint imposed\cite{Gebhard1990,Bunemann2012}
\begin{equation}
 \hat{P}_{il}^2\equiv 1+x_{il}\hat{d}^{\textrm{HF}}_{il}\;,
 \label{eq:condition}
\end{equation}
where $\hat{d}^{\textrm{HF}}_{il}=\hat{n}_{il\uparrow}^{\textrm{HF}}\hat{n}_{il\downarrow}^{\textrm{HF}}$, $\hat{n}_{il\sigma}^{\textrm{HF}}=\hat{n}_{il\sigma}-\langle\hat{n}_{il\sigma}\rangle_0$. In such approach one can express all $\lambda_{\Gamma|il}$ parameters from Eq. (\ref{eq:P_Gamma}) by using $x_{il}$
\begin{equation}
\begin{split}
    \lambda^2_{dl}&=1+x_l(1-\langle\hat{n}_{il\sigma}\rangle_0)^2,\\
    \lambda^2_{sl}&=1-x_l\langle\hat{n}_{il\sigma}\rangle_0(1-\langle\hat{n}_{il\sigma}\rangle_0),\\
    \lambda^2_{\emptyset l}&=1+x_l\langle\hat{n}_{il\sigma}\rangle_0^2,\\
\end{split}
\end{equation}
where $\langle...\rangle$ denotes the expectation value in the non-correlated state $|\Psi_0\rangle$ and we assume that the we are considering a homogeneous situation with no magnetic ordering $\lambda_{\sigma|il}\equiv\lambda_{sl}$, $\lambda_{\uparrow\downarrow|il}\equiv\lambda_{dl}$, $\lambda_{\emptyset|il}\equiv\lambda_{\emptyset l}$, and $x_{il}\equiv x_{l}$. In fact, it is convenient to treat $x_{l}$ as the variational parameters instead of $\lambda_{\Gamma|il}$. Approach based on the constraint (\ref{eq:condition}) allows to significantly improve the Gutzwiller approximation and obtain the full Gutzwiller wave function solution in the higher orders of the so-called diagrammatic expansion of the Gutzwiller wave function (DE-GWF)\cite{Zegrodnik2018,Zegrodnik2019}. However, for the purposes of this analysis the the zeroth order expansion, expressed in Eq. (\ref{eq:H_SGA}), is sufficient enough and is equivalent to the Gutzwiller approximation. The latter allows to cast $\langle H\rangle_G=\langle\Psi_G|\hat{H}|\Psi_G\rangle/\langle\Psi_G|\Psi_G\rangle$ in a relatively compact form. Below we show the expressions for the expectation values in the correlated state from all the three terms contributing to the system energy [cf. Eq. (\ref{eq:Hamiltonian_general})]
\begin{equation}
    \begin{split}
        \langle\hat{H}_{TBA}\rangle_G&=\sum_{\langle ijll'\rangle}q_l q_{l'}t^{ll'}_{ij}\langle\hat{c}^{\dagger}_{il\sigma}\hat{c}_{jl'\sigma} \rangle_0\\
        &+t_0\sum_{il}\bigg(\lambda^2_{sl}\langle\hat{n}_{il\sigma}\rangle_0+(\lambda^2_{dl}-\lambda^2_{sl})\langle\hat{n}_{il\sigma}\rangle_0^2\bigg),
    \end{split}
    \label{eq:H_TBA_SGA}
\end{equation}

\begin{equation}
    \begin{split}
        \langle\hat{H}_U\rangle_G&=U\sum_{il}\lambda^2_{dl}\langle\hat{n}_{il\sigma}\rangle_0^2+V\sideset{}{'}\sum_{ill'\sigma\sigma'}\bigg(4\langle\hat{n}_{il\sigma}\rangle_0\langle\hat{n}_{il'\sigma'}\rangle_0\\
        &+2(2-\lambda_{sl}^2)(2-\lambda_{sl'}^2)\langle\hat{c}^{\dagger}_{il\sigma}\hat{c}_{il'\sigma}\rangle_0\bigg),
    \end{split}
    \label{eq:H_U_SGA}
\end{equation}

\begin{equation}
    \begin{split}
        \langle\hat{H}_{SC}\rangle_G&=-J\sum_{ijl}\lambda_{sl}^4\langle\hat{c}^{\dagger}_{il\uparrow}\hat{c}^{\dagger}_{jl\downarrow}\hat{c}_{il\downarrow}\hat{c}_{jl\uparrow}\rangle_0\\
        &-J'\sideset{}{'}\sum_{ijll'}\lambda_{sl}^2\lambda_{sl'}^2\langle\hat{c}^{\dagger}_{il\uparrow}\hat{c}^{\dagger}_{jl\downarrow}\hat{c}_{il'\downarrow}\hat{c}_{jl'\uparrow}\rangle_0,
    \end{split}
    \label{eq:H_SGA}
\end{equation}
where
\begin{equation}
    q_l\equiv\lambda_{sl}\bigg(\lambda_{dl}\langle\hat{n}_{il\sigma}\rangle_0+\lambda_{\emptyset l}(1-\langle\hat{n}_{il\sigma}\rangle_0)\bigg).
\end{equation}
As one can see the expectation value of system energy in the correlated state is expressed in terms of the expectation values in the noncorrelated state but premultiplied by proper renormalization factors, which are dependant on the variational parameters and average number of particles in particular local states. 

It has been shown that in order to ensure the statistical consistency condition during the energy minimization procedure one needs to supply the expression for $\langle\hat{H}\rangle_G$ with the Lagrange-multiplier terms leading to the auxiliary energy operator of the form\cite{Jedrak2011}
\begin{equation}
\begin{split}
    \hat{K}&=\langle\hat{H}\rangle_G+\sum_{\langle ijll'\rangle}\tilde{t}^{ll'}_{ij}\bigg(\hat{c}^{\dagger}_{il\sigma}\hat{c}_{jl'\sigma}-\langle\hat{c}^{\dagger}_{il\sigma}\hat{c}_{jl'\sigma}\rangle_0\bigg)\\
    &+\sum_{\langle ijl\rangle}\bigg[\tilde{\Delta}_{ijl}\bigg(\hat{c}^{\dagger}_{il\uparrow}\hat{c}^{\dagger}_{jl\downarrow}-\langle\hat{c}^{\dagger}_{il\uparrow}\hat{c}^{\dagger}_{jl\downarrow}\rangle_0\bigg)+H.c.\bigg],
\end{split}
\end{equation}
where $\tilde{t}^{ll'}_{ij}$, $\tilde{\Delta}_{ijl}$ are the Lagrange-multipliers. Summation in the first term of the above is carried out over the same sites as the corresponding summation in the electron-hopping term of $\hat{H}_{TBA}$. The second summation is carried out over the nearest neighbors only since only such pairing amplitudes are taken into account here.

The final step of the procedure is the minimization of the free energy potential corresponding to the auxiliary energy operator $\hat{K}$ over the mean-field hopping and pairing expectation values $\langle\hat{c}^{\dagger}_{il\sigma}\hat{c}_{jl'\sigma}\rangle_0$, $\langle\hat{c}^{\dagger}_{il\uparrow}\hat{c}^{\dagger}_{jl\downarrow}\rangle_0$, the respective Lagrange multipliers $\tilde t^{ll'}_{ij}$, $\tilde{\Delta}^{ll'}_{ij}$, as well as the variational parameters $x_{l}$. Such procedure constitutes the so-called Statistically consistent Gutzwiller approximation (SGA). For mode details of the method itself see Refs. \onlinecite{Jedrak2011,Kaczmarczyk2013,Zegrodnik_2013}.

\bibliography{refs.bib}

\end{document}